# Shape of Empirical and Synthetic Isoseismals:
## Comparison for Italian M ≤ 6 Earthquakes

G.M. Molchan[1,2], T.L. Kronrod[1,2], G.F. Panza[2,3]

[1] *International Institute of Earthquake Prediction Theory and Mathematical Geophysics, Russian Academy of Sciences, Profsoyuznaya., 84/32, Moscow 117 997 Russia*
[2] *The Abdus Salam International Center for Theoretical Physics, SAND Group, Trieste, Italy*
[3] *Dipartimento di Scienze della Terra, Trieste University, Trieste, Italy*

*Abstract.* We present results from a comparative analysis of empirical and synthetic shapes for isoseismals of low intensity ($I$ = IV−VI on the MCS scale) for six Italian earthquakes of ML = 4.5−6. Our modelling of isoseismals is based on a plane-stratified earth model and on the double−couple point source approximation to calculate seismograms in the frequency range $f \leq 1$ Hz. With a minimum of parameter variation we demonstrate that the low intensity isoseismals provide information on source geometry. We strive to avoid subjectivity in isoseismal constructions by using the new Diffuse Boundary method, which visualizes isoseismals with their uncertainty. Similar results in this direction are known for large earthquakes (ML ≈ 6 or greater) with extended sources and for the higher isoseismals ($I \geq$ VI on the MM scale). The latter studies disregard the earth structure, use a greater number of parameters, and therefore have greater possibilities for fitting the data than our approach.

## 1. Introduction

Seismic hazard analyses deal with macroseismic intensity $I$ and/or peak ground acceleration $a_p$. Acceleration is usually preferred, although the two quantities are essentially different. Macroseismic intensity is a measure of the *effect* that seismic motion produces on man and structures. This is influenced by resonance and residual displacements, the latter being only indirectly related to acceleration amplitude. The estimates of $I$ are based on expert evaluations and questionnaire surveys, and, in this respect, intensity is less preferred when compared with accurate instrumental measurements of $a_p$. However, while $a_p$ is measured at a site, $I$ is relevant for an area, i.e., a group of type structures or the perceptions of people in a town or village as a whole. It follows that $I$ has a statistical nature, so it is more stable than single−site $a_p$ measurements. $I$–data are usually more dense in space. Furthermore, in spite of the Gutenberg−Richter law, the number of strong motion records with $I \leq 5$ (the MM scale) does not increase with decreasing $I$, because the engineers are not interested in low intensity excitations (see, e.g., Aptikaev, 2001).

Theoretical modeling is only feasible for $a_p$, since it is extremely difficult to reproduce the detailed effects of an earthquake. The reason lies in the fact that the types of structures are rather diverse and are subject to aging, and the soil water content is always changing. Consequently, every observed $I$–data set is the result of a unique natural experiment, so it is not without reason that in recent years increasing attention is being paid to the development of macroseismic data bases (see Boschi et al., 1997, 2000; Monachesi and Stucchi, 1997).

Instrumental observations of $a_p$ are primarily needed to set up building codes for seismic regions. With this goal in view, probability maps of expected ground motion acceleration are developed. However, when the insurance of type structures is at issue, one would prefer to use analogous maps of intensity, since intensity is directly related to damage statistics. It thus appears that none of the two quantities, $I$ and $a_p$, is the one to be preferred. They are different and complementary quantities to be used jointly in seismic risk analysis.

$I$–data is successfully used to parameterize historical earthquakes, to be more specific, to determine magnitude and hypocenter position, as well as source azimuth and length for large events. The recent publications on this topic include Bakun and Wentworth(1997); Cecić et al. (1996); Gasperini et al. (1999); Gasperini (2001); Johnston (1999); Musson (1996); Peruzza (2000). The remaining geometric source parameters (dip and rake) may be reflected in the shape of isoseismals as suggested in the paper by Panza et al.(1991) (the history of relevant research can also be found in this reference), that seems to have been the first study to compare empirical isoseismals with theoretical ones, derived from complete synthetic seismograms for realistic earth models. The comparison was purely qualita-tive, since the theoretical isoseismals had not been calibrated in terms of intensity, while the empirical isoseismals provided no indication of their accuracy.

A characteristic frequency range (f = 0.5–10 Hz) is associated with macroseismic intensity $I$ = IV–X (MSK scale) (Sokolov and Chernov, 1998). Consequently, the high frequency approximation of seismograms developed by Bernard and Madariaga (1984) and Spudich and Fraser (1984) is of interest for the comparison between $a_p$ and $I$. This approximation has been used recently as a theoretical background for a simple model of Intensity Attenuation Law for large earthquakes (Berardi et al., 1995; Sirovich, 1996; Pettenati et al., 1999). In particular, the Attenuation Law developed by Sirovich (1996) incorporates the direction of rupture propagation and the radiation pattern due to fault plane geometry. Since crustal inhomogeneity is disregarded, these generalizations of the $I$–Attenuation are meant to deal with the geometry of the higher isoseismals for large ($M_w > 5.7$) events (Sirovich, 1996). It was shown that the new $I$–Attenuation models are good in some cases for fitting the higher isoseismals, especially when this fitting includes an individual calibration of $a_p$ (in terms of $I$) for each event. Thanks to the simplicity of the calculation involved, these approaches can possibly be used to invert $I$–data into source parameters. The question that arises however is, when the inversion is feasible and what are the bias and scatter due to unaccounted–for earth inhomogeneity; furthermore the interpretation of "low–level experimental isoseismal shapes" remains a "highly questionable" problem (Sirovich, 1996).

The present study proceeds on the lines of Panza et al. (1991). The complete synthetic seismograms are computed in the frequency range ≤ 1 Hz assuming lateral homogeneity of the crust and the point source model. For this reason we analyze the lower isoseismals of moderate ($5 \leq M_L \leq 6$) earthquakes, i.e., consider a situation that is adjacent to that described above as to M and $I$. The point source model involves the least possible number of source parameters, while the synthetic fields are calibrated in the same manner for all events. These limitations seriously constrain the degrees of freedom available for fitting the $I$–data.

The new part of our approach consists in mapping the isoseismal uncertainty. In recent years several formalized algorithms for isoseismal tracing have been put forward (Tosi et al., 1995; Molchan et al., 2002; Sirovich et al., 2002), and a persistent skepticism simultaneously arose regarding the use of isoseismals in the $I$–data analysis (see, e.g., Gasperini et al., 1999; Pettenati et al., 1999). The skepticism stems from the fact that the formalization of isoseismals taken by itself cannot overcome the subjectivity in isoseismal tracing, because the procedure invariably involves adjustable parameters. Speaking the language of mathematical statistics, one could restate the argument as follows: a point estimate of a parameter has little meaning without an associated confidence interval. For this reason we need a substitute interval estimation for the case of isoseismals as well, resulting in diffuse isoseismal boundaries. The problem has been treated by Molchan et al. (2002), while Kronrod et al . (2002) have presented ordinary- and diffuse-boundary isoseismals of good quality for 55 earthquakes from Italian $I$–data bases (Boschi et al., 1997, 2000; and Monachesi and Stucchi, 1997). The present study aims to demonstrate, with examples of moderate events taken from the atlas of Kronrod et al. (2002), that diffuse boundaries provide some information on the source geometry.

## 2. Synthetic Intensity

We use the modal summation technique for Rayleigh and Love waves (Panza, 1985; Florsch et al., 1991) to compute seismograms to be recorded within 200 km of the epicenter in the frequency range f ≤ 1 Hz. The technique assumes the double-couple source model and a horizontally stratified earth, thusseverely limits the degrees of freedom in the problem of comparing theoretical and empirical intensities.

*Theoretical Intensity.* There are many characteristics of strong motion that correlate with $I$: peak displacement, peak velocity, and peak acceleration; the Arias intensity (Arias, 1970); Significant Acceleration (Bolt and Abrahamson, 1982) and some others. It is not however entirely clear which of these correlates with $I$ best, and the problem seems to admit no unique solution. According to Sokolov and Chernov (1998) and Chernov and Sokolov (1999), each level of intensity $I$ has its own optimal frequency band, which explains the diversity of instrumental counterparts of $I$. It is also known that, having to select between peak velocity and peak acceleration for the seismic design range $I > 6$ (MM scale), one should prefer the former (Ambraseys, 1974; Aptikaev, 2001). However, since we use the point source model, we shall be interested in the interval $I$ = V–VII (MCS scale), which is, as a matter of fact, adjacent to $I > 6$ (MM scale).

The choice of the theoretical counterpart of *I* and the calibration of it are interrelated problems. Dealing with synthetic seismograms for the range $f \leq 1$ Hz we have suitable data for calibrating the above peak quantities only (see below). Our final option is the peak acceleration at frequencies $f \leq 1$ Hz, $\hat{a}_p$, as the theoretical counterpart of *I*. Since the vertical peak acceleration is usually smaller than the horizontal and since building are mainly damaged by horizontal motion, $\hat{a}_p$ was taken to be $\max_t \|a^H(t)\|$, where $a^H$ is the horizontal acceleration vector and $\|\cdot\|$ is its norm.

*Source Parameters.* The point source model involves the following parameters: epicenter, depth ($h$), moment magnitude $M_w$, and the angles that specify the focal mechanism: $(\psi, \delta, \lambda) =$ (azimuth, dip, rake) or FPS (Fault Plane Solution) for brevity. This is the minimal possible set of source parameters. According to Gusev and Shumilina (2000), the point source model is valid at distances greater than 1.5 the source length, nevertheless to be conservative we will consider, in what follows, comparatively small events ($M_w \leq 6$) and examine lower empirical isoseismals.

Modeling of an event where the source is actually extended by using the point source model generates an uncertainty in *h* that is of the order of the source length. The typical magnitude uncertainty is 0.3, while an uncertainty of $\leq 20°$ is commonly assumed for the focal angles in cases where the fault plane solution is based on P−wave first motion data and is classified to belong to the highest category, class A (Frepoli and Amato, 1997).

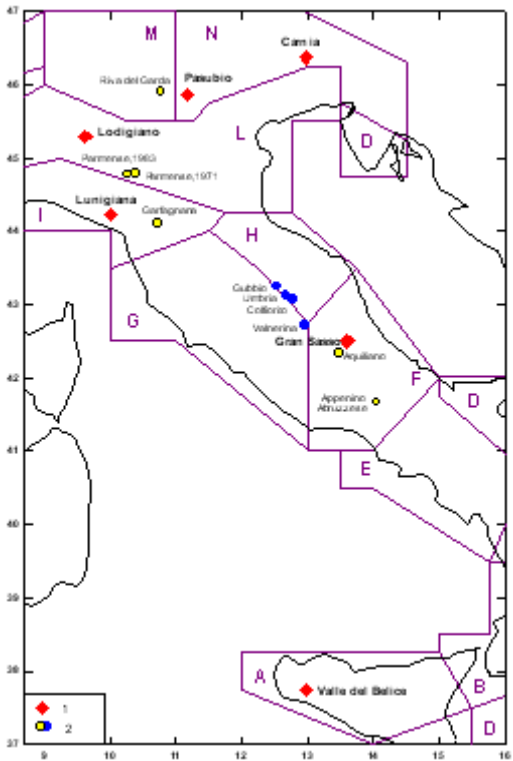

**Fig. 1**. Structural zones of Italy (**A** through **N**) and events discussed in the text.
*Notations*: 1 – successfully modelled events (see Fig. 2–7); 2 – other events involved in the comparison of empirical and synthetic data (24/06/1958, Aquiliano, $M_L = 5.0$; 15/07/1971, Parmense, $M_L = 5.3$; 13/12/1976, Riva del Garda, $M_L = 4.4$; 09/19/1979, Valnerina, $M_L = 5.5$; 09/11/1983, Parmense, $M_L = 4.9$; 29/04/1984, Gubbio, $M_L = 5.0$; 07/05/1984, Appenino Abruzzese, $m_b = 5.4$; 05/06/1993, Umbria, $M_L = 4.1$; 24/08/1995, Garfagnana, $M_d = 4.2$; 26/09/1997, Colfiorito, $M_w = 6.0$).

A horizontally stratified earth is specified by velocity functions of depth for P and S waves, $V_P(h)$ and $V_S(h)$, the respective quality factors $Q_P(h)$ and $Q_S(h)$, and a density function $D(h)$. The earth model used for Italy (Costa et al., 1993) is based on data found in the literature. The velocity parameters are not uniform throughout the lithosphere in Italy, so Costa et al. (1993) proposed a crude division of the lithosphere into several structural zones (Fig. 1), each having its own D, V and Q. The scale of lateral averaging for these functions must exceed the size of the isoseismals of interest. We recall that the $\hat{a}_p$ field is calculated for a horizontally stratified earth. For this reason $\hat{a}_p$ may involve appreciable discontinuities across zone boundaries, when the parameters in the adjacent zones differ by significant amounts. Therefore, the presence of different structural zones means that any comparison of *I* and $\hat{a}_p$ fields must largely be based on those isoseismals which nearly fit into a single structural zone. We say "nearly", because zone boundaries can well be transition zones in their own right.

The modal summation technique used here to compute seismograms is applied to events with a fixed moment, $M_0 = 1 \cdot 10^{-7}$ Nm. For an arbitrary moment the seismic signal spectrum is rescaled (differently at different frequencies) using the Gusev source spectrum (Gusev, 1983; cf. an updated version in (Aki, 1987)). The spectrum was derived by averaging worldwide data and is, according to (Gusev, 1983), a compromise to smooth both intra- and interregional variations. In practice this may produce a bias in the $a_p$ ampli-Itude, i.e., a mismatch between $a_p$ and $I$, in cases where the $a_p - I$ relation is specified beforehand for the region of study. *The Calibration of* $\hat{a}_p$. There are numerous empirical relations between peak acceleration and intensity. However, no $I / \hat{a}_p$ relation is available with the single exception of Panza et al. (1999). These authors were comparing two types of field. The one, $I_{max}(g)$, is the highest intensity ever observed at a point g. The other, $\hat{a}_p^{max}(g)$, is a hypothetical maximum peak acceleration at point g in the frequency range $f \leq 1$ Hz. The map of $\hat{a}_p^{max}$ is obtained by computing a family of accelerograms using earth models and focal mechanisms typical of Italian areas (Costa et al., 1993). The epicenters of the hypothetical events fill a regular grid, the depth of each event being assumed to be equal to 10 km for $M_w < 7$ and 15 km for $M_w \geq 7$. The magnitude of an event is set equal to the hypothetical maximum magnitude, $M_w^{max}$, for the site of interest. The next step is to find, for each point on the map, the maximum value of the peak acceleration, $\hat{a}_p^{max}$, that has been produced by some event of the family. Panza et al. (1999) compared $I_{max}$ and $\hat{a}_p^{max}$ and obtained the regression equation

$$\log \hat{a}_p \, [\text{cm/sec}^2] = b_0 + b_1 I \pm \varepsilon \tag{1}$$

where $\varepsilon$ is the regression residual, $b_0 = -1.61$ and $b_1 = 0.35$.

Since the models and the methodology we are using to compute seismograms are the same as in (Panza et al., 1999), we quantize the $\hat{a}_p$ scale according to the relation

$$\hat{a}_p = 3 \cdot 2^{(I_a - 6)} \, [\text{cm/sec}^2], \tag{2}$$

where the integer quantity $I_a$ will be called the theoretical (synthetic) intensity. Relation (2) means that we have adopted the rule of doubling $\hat{a}_p$ for a change of one in $I_a$, and we have $\hat{a}_p = 3$ for $I_a$ = VII according to (1).

Relation (2) is a very crude one. For instance, Panza et al. (1999) note that $\hat{a}_p$ shows a better correlation with two variables, intensity and distance. Besides, regression (1) we are using was derived for fixed depths of the hypothetical events, and this assumption may have affected $b_0$. Therefore, the relation between macroseismic intensity $I$ and its synthetic counterpart, whenever the latter is available, cannot possibly be an exact one. If the isolines of $\hat{a}_p$ are assumed to be similar for adjacent levels, the expected invariants in the $I$ vs. $\hat{a}_p$ comparison could be the orientation and shape of the isolines. The variation of the inexact parameters $M_w$ and $h$ is a suitable tool for adjusting the theoretical isoseismal area.

### 3. Isoseismals and the Visualization of Isoseismal Uncertainty

The intensity model does not incorporate site effects due to small–scale geologic and topographic heterogeneities. This difficulty can be overcome to a certain extent by generalizing $I$–data as isoseismals which act as a smoothing filter. Molchan et al. (2002) have developed two methods for tracing isoseismals. The one is a smoothing technique, while the other also visualizes the uncertainty of an isoseismal. Both techniques examine the shape of isoseismal areas, and for this reason play a leading role in solving the problem of comparing spatial distributions of $I$ and $\hat{a}_p$. In order to make this paper self–contained, we recapitulate briefly the two techniques.

*The Modified Polynomial Filtering Method (MPF Method)*

The method mainly aims at reducing the noise component in the data, including small–scale site effects. A circle $B_R(g)$ of radius R is centered at a point g of a regular grid. The radius is chosen so that at least $n_p(g_k, I_k)$ points fall into the circle. The data in the circle are fitted with a surface of degree 2, $P_2(.)$, by the method of least squares. The value of $P_2(.)$ at the center, g, is assumed to be the estimate $\hat{I}(g)$ of $I$ at g. Since $I$ is discrete, the radius R can be increased in the interval $(0, R_0)$ until the number of different intensity values in $B_R(g)$ exceeds a specified amount $n_I$. The greater the data noise, the higher is $n_I$. The introduction of the parameter $n_I$ allows the highest degree of smoothing for the data in the annulus between adjacent isolines, where the field ought to be constant by assumption. The typical values of $n_I$ and $R_0$ for Italy are $n_I = 3$, $R_0 = min(70 \text{ km}, D/4)$, where D is the diameter of the $\{g_k\}$ points. The $I$ distribution at the periphery is estimated at a point g, when the points that fall into $B_R(g)$ are seen at an angle $\varphi > 200^o$ looking from the center, thus avoiding unjustified extrapolation of the $I$ observations. For the resulting smoothed $\hat{I}$ (g) field, the area where $\hat{I}$ (g) > (I – Δ/2) is adopted as the isoseismal area of level *I*. Here Δ is the step in *I*.

*The Diffuse Boundary Method* (DB method).

The idea of this method is more easily understood, if we consider the one–dimensional case of ($g_k$, $I_k$) observations. Suppose for the moment that isoseismals are embedded and that it is required to divide $I_k ≥ J$ points (labelled '+' here) and $I_k < J$ points (labelled '–') which lie on a line. In that case there will be (in the absence of noise) a cluster of pluses enclosed within two clusters of minuses. The empty intervals $Δ_1$ and $Δ_2$ that divide the pluses and the minuses cover the true boundary points. They are taken to be the diffuse isoseismal boundary of level J. Only some additional information and not the choice of the method can reduce the objective uncertainty of the boundary $Δ_1 \cup Δ_2$.

Let us suppose that the data involve some low noise. The cluster of pluses will then be subdivided, by '–' points, into a series of smaller clusters. Below we show schematically a variant where the plot starts from the barycenter of the pluses taken at zero, and it is required to find an analogue of the interval boundary $Δ_2$.

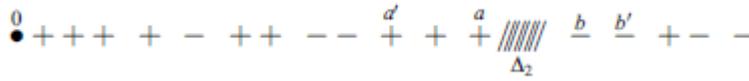

We specify a small parameter ε, to embody our notion of the noise level that is present in the data. It would be natural to suppose that pluses surrounded by minuses, when found at the periphery (on the right in our figure), must be erroneous. Consequently, we will find an interval $Δ_2$ = (a, b) that separates the cluster of pluses [a', a] and the cluster of minuses [b, b'] and which has the following property: the number of pluses in the interval (a, ∞) is less than ε·100% of the total number of pluses in (0, ∞), while the number of pluses in [a', ∞) is greater than ε·100%. In this case $Δ_2$ is taken to be the right–hand diffuse boundary between pluses and minuses. If the set of pluses is not a connected one, then $Δ_2$ is the interval estimate of the right-hand boundary itself. A similar definition is valid for $Δ_1$. In this way the parameter ε specifies the threshold of the possible error in the peripheral pluses.

The two–dimensional case can be reduced to solving a family of one–dimensional problems as follows. Let us assume that the isoseismal area $G_I$, where the intensity is greater than or equal to *I* is convex. Let us take in turn all "*l*" lines (as a matter of fact, this can be done at some discrete interval in the space of their parameters) that intersect the area which contains the intensity points. The {$g_k$} points from the H–neighborhood of an *l* line are projected onto *l*, the next step being to solve the above one–dimensional problem for *l* and *I*. The result will be two intervals $Δ_1$ and $Δ_2$ on the *l* line which characterize the uncertainty of an isoseismal of level *I*, when the isoseismal is viewed along the *l* ray. The totality of all intervals forms a jagged diffuse isoseismal boundary of level *I* (to be called the DB isoseismal from now on). The boundary looks rather unconventional, being as it is a family of line segments of various lengths and all possible directions. The decision about a diffuse boundary with respect to an individual straight line is not stable. However, a population of these decisions yields an additional quantity, namely, the intensity of superposition of the intervals, $Δ_i$ (*l*), which makes the boundary rather stable.

The algorithm described above is valid for a convex $G_I$. As a matter of fact, $G_I$ is not convex and it is not always simply connected, even when site effects have been taken into account. Consequently, a DB isoseismal will lose some of the $Δ_i$ intervals for those boundary points of $G_I$ which are internal to the convex hull of $G_I$. This is due to the fact that only two (leftmost and rightmost) boundary points of $G_I$ are taken into consideration in any *I*–data cross–section. Nevertheless, this does not impede the identification of large–scale disconnected components of $G_I$ (see Kronrod et al., 2002; Molchan et al., 2002) or of peculiarities of the boundary of $G_I$. For example, a well–defined cross shape of the *I* = V isoseismal is seen in Fig. 2b instead of the conventional oval. Some elements of a cross shape can tentatively be discerned in the *I* = IV isoseismal derived by the MPF method (Fig. 2a).

The DB method involves two basic parameters: the bandwidth H and the noise parameter ε. The former is a smoothing parameter governed by the density of the observations: the higher the density, the smaller is H. The typical ε – value for Italy is 5%, while H is 20–40 km.

We conclude by noting that the DB method is well adapted to deal with the comparison of $a_p$ and *I* distribution in space. The preceding analysis shows that the relation between $I_a$ and *I* cannot be an exact one, while a diffuse boundary can well estimate the shape of an isoseismal, but it is a rather poor estimate of its area.

## 4. Comparison of Synthetic and Empirical Intensities

Our joint analysis of $I$ (MCS scale) and $I_a$ is based on sixteen earthquakes of magnitude $M_L \leq 6$ (Fig. 1). These sixteen are those events in the Atlas of Kronrod et al. (2002) for which the equivalent point source parameters are available, both MPF and DB techniques yield satisfactory isoseismals, and the number of site intensity observations is ~100 or greater. The isoseismals for each ($I$, $I_a$) pair are obtained using all $I$−data points and identical parameter sets. The comparison involves diffuse boundaries, the MPF method being merely used to identify strong local anomalies in the $I$−data. We test our conclusions about the relation between $I$ and $I_a$ by adding residuals at anomalous points to $I_a$. When the conclusion remained unchanged, we treated it as robust with respect to the noise component in the $I$−data. (Part of the noise could also be caused by small-scale site effects among other factors).

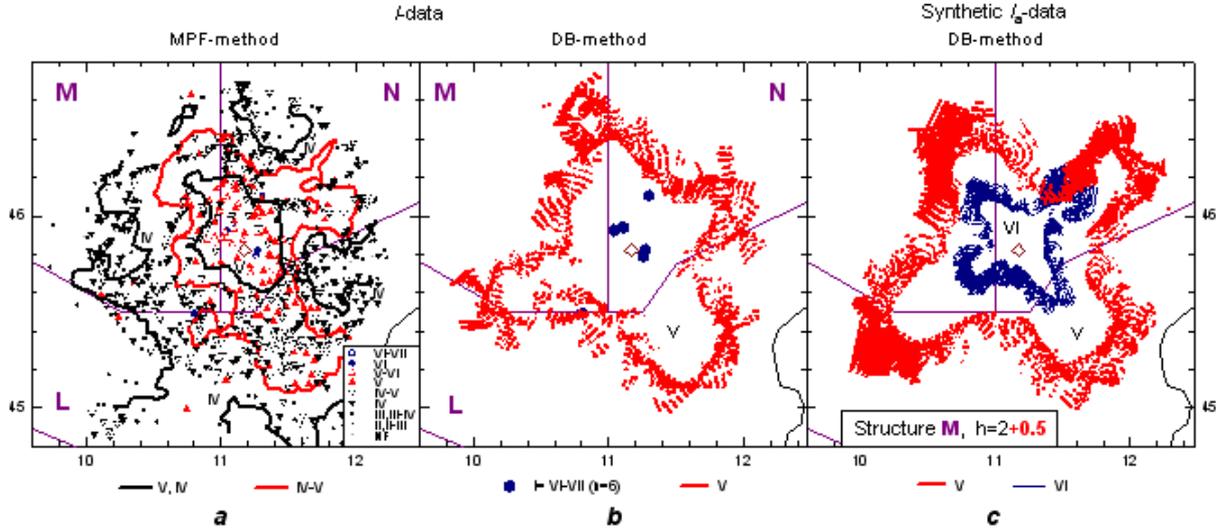

**Fig. 2.** 13/09/1989 Pasubio earthquake:
epicenter (45.86, 11.17); $h$ = 2 (CS, 2001); FPS = (145°,85°,-180°) by A. Frepoli (class A); $M_L$ = 5.1 (ISC), 4.4 (CS, 2001)
a) MPF isoseismals and observations (symbols); b) $I$ = VI–VII raw data (symbols) and DB isoseismal of level $I$ = V; c) theoretical DB isoseismals of levels $I_a$ = V, VI. Background: epicenter (rhombus) and structural zones **M**, **N**, **L**.

We use the trial−and−error strategy in our comparison between $I$ and $I_a$ fields. The source parameters (depth, magnitude, and FPS) borrowed from the literature are treated as the basic ones. In the case of multivalence of a parameter all its values are basic for us unless more arguments are advanced. In our strategy the relevant standard deviation of a parameter is the guide to fit $I$−data. The fitting possibilities have been extremely restricted owing to the time factor. A systematic variation of values of two or three parameters was not feasible. For this reason our fitting is far from optimal.

Below we discuss six events for which the synthetic and empirical DB isoseismals are similar enough. The comparison between $I$ and $I_a$ involves individual isoseismals that are mostly within a single structural zone or within zones with similar parameters (Fig. 1). Our discussion of the six events will be confined to those parameters which have been varied in the analysis. They are specified in the figures either by giving the name of a structure (when the isoseismal intersects more structures than that specified) or as $a + \delta$, where $a$ is the value of a parameter taken from the literature and $\delta$ is our correction with the sign.

**The earthquake of 13/09/1989**, $M_L$ = 4.4 (CS), 5.1 (ISC), $I_0$ = VI–VII, **Pasubio** (foothills of the Dolomite Alps). The $I$ map is from BM (1989), the number of observations is $N_{obs}$ = 779, and the results of the application of the MPF and DB methods are shown in Fig. 2.

The DB method when applied to these data identifies the $I$ = V isoseismal pretty clearly. One observation site (Suzzara) has been eliminated from the data in Fig. 2b, the reason being that the intensity value there seems to have been overestimated by one intensity unit. From Fig. 2b it is seen that the DB isoseismal of level $I$ = V is cross−shaped. This kind of seismic energy radiation is typical of pure strike slip earthquakes. We asked A. Frepoli to determine the earthquake mechanism for this event, and his results confirm our strike slip hypothesis: dip δ = 85°, rake λ = -180°. A more accurate inference can be drawn by comparing the $I$=V isoseismal to its theoretical counterpart, the isoline of $I_a$ = V (Fig. 2c). Most of the $I$=V isoseismal occupies two structural zones, **M** and **N** (see Fig. 2), just a small part falling into zone **L**. The earth models for **M** and **N** are very similar, so the theoretical calculations of $â_p$ were based on the parameters of a single structural zone **M**, and the subsequent conclusions are not affected by the choice of **M** or **N**.

The similarity of the DB isoseismals of level V is achieved by using the magnitude $M_w$ = 5.1 and the depth $h$ = 2.5 km instead of $h$ = 2 km (CS, 2001). The choice $M_w$ = 5.1 instead of $M_L$ = 4.4 (CS, 2001) is in agreement with the values given by the International Seismological Centre, namely, $M_L$ = 5.1 and $m_b$ = 5.1 (NAO Network). Besides, Peresan et al. (2000) note that the post–1987 $M_L$ magnitudes for Italian earthquakes, as reported in the ING bulletins, are underestimated by an average of 0.5. For this reason the estimate $M_L$ = 4.4 is less preferable.

The empirical $I$–data for this earthquake are contaminated by a strong "noise". For example, the residual of $I$ is found to be in excess of one unit, after smoothing the data by the MPF method, at 42 sites (of a total of 779). Adding these residuals to the theoretical values of $I_a$ does not affect the shape of the theoretical DB isoseismal at $I_a$ = V. In other words, the conclusion that the $I$ and $I_a$ isoseismals of level V are similar will not be affected by the noise component that may be present in the $I$–data

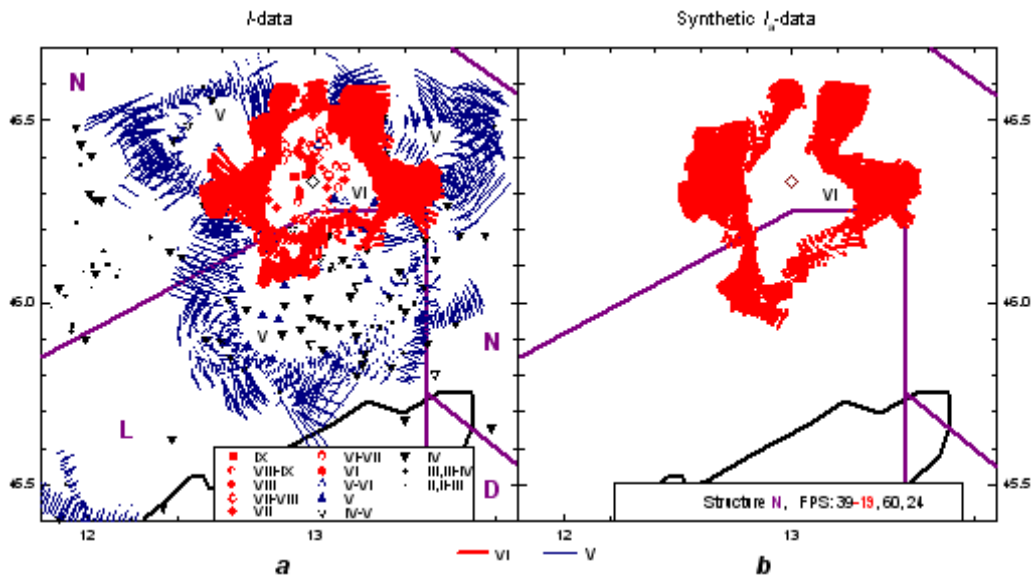

**Fig. 3.** 27/03/1928 Carnia earthquake: epicenter (46.38, 12.98); $h$ = 4 (Kunze, 1982), 5 (NT, 1997), 20 (Cagnetti et al., 1976); FPS = (39°,60°,24°) (Cagnetti et al., 1976), (112°,90°,0°) (GI, 1985); $M_L$ = 5.6 (NT, 1997), 5.8 (Cagnetti et al., 1976); a) raw data (symbols) and DB isoseismals of levels $I$ = VI, V; b) theoretical DB isoseismal of level $I_a$ = VI. Background: epicenter (rhombus) and structural zones **N**, **L**.

***The earthquake of 27/03/1928***, $M_L$ = 5.6, $I_0$ = VIII, ***Carnia*** (Carnian Alps). The $I$ map is from Boschi et al. (2000), $N_{obs}$ = 289, and the result of the application of the DB method is shown in Fig. 3.

Figure 3 compares the DB isoseismals of level VI for $I$ and $I_a$. The $I$>V isoseismal area is entirely contained in the structural zone **N** (Fig. 1), while, to the south, it approaches the boundary of zone **L**, representative of the Padan basin, where the structural parameters are significantly different from those of **L**. For this reason it is unlikely that the details in the southern part of the $I$ = VI isoseismal can be identified. In the example we are discussing, we choose the value $h$ = 4 among the three available estimates $h$ = 4.5 and 20 km and we set $M_w$ = $M_L$ = 5.6. Only the azimuth of the fault has been modified by (-19°), i.e., the angles (20°, 60°, 24°) have been used instead of the ones given by the FPS=(39°, 60°, 24°). The FPS solution can hardly belong to class A, since there is a different option in the literature (GI, 1985): FPS=(112°, 90°, 0°) which shows that the strike slip nodal planes have been determined inaccurately. Nevertheless, our correction to the azimuth does not exceed 20°. Overall, the DB boundaries of level VI for the observed and theoretical intensities are in a qualitative agreement.

***The earthquake of 15/05/1951***, $M_L = 5.0$, $I_0 =$ VI, ***Lodigiano*** (western Padan basin, Lombardy). The *I* map is from Monachesi and Stucchi, 1997), $N_{obs} = 88$ for the main shock and $N_{obs} = 32$ for the aftershock, and the result of the application of the DB method is shown in Fig. 4.

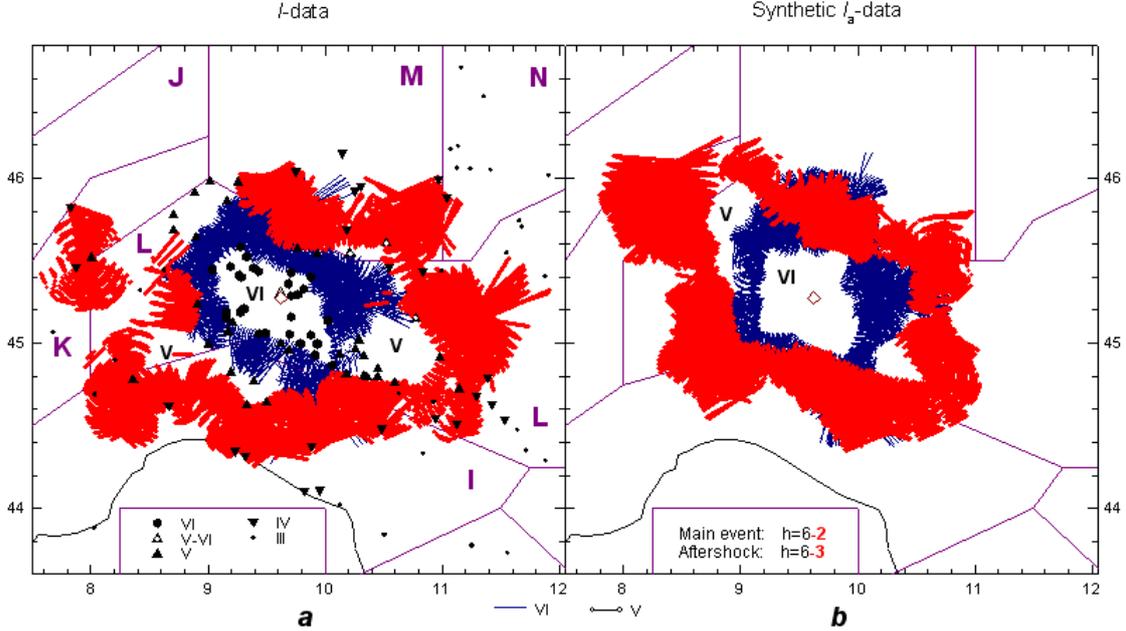

**Fig. 4**. 15/05/1951 Lodigiano earthquake: epicenter (45.30, 9.62); $h = 6$ (GI, 1985), 12 (NT, 1997); FPS = (236°,74°,192°) (GI, 1985); $M_L = 4.9$ (NT, 1997), 5.0 (GI, 1985), 5.5 (NEIC). Aftershock (epicenter: triangle): $h = 6$ (GI, 1985), FPS = (221°,74°,209°) (GI, 1985), M = 4.5 (GI, 1985); a) raw data (symbols) and DB isoseismals of level *I* = VI, V; b) theoretical DB isoseismals of level $I_a$ = V, VI. Background: epicenter (rhombus) and structural zones **I**, **J**, **K**, **L**, **M**, **N**.

The event is not easy to analyze, because the I V isoseismal zone (rank 1) is in the Padan basin, while its boundary is in the transition zones between M and L to the north, and between I and L to the south, the structural parameters for M and I being very different from those for L. For this reason the âp field for each of the three zones (L, M, N) was calculated from the structural parameters of its own zone. The main shock, ML = 5.0, h =6 km, FPS = (221°, 74°, 209°), was followed by an ML = 4.2–4.6 aftershock. When one deals with a sequence of large events, the counterpart of the theoretical intensity I a at a site is the largest of the theoretical values of I a corresponding to these events. In the example under discussion, the value M = 5.0 was selected from the three magnitude estimates ML = 4.9, 5.0, 5.5 available for the main shock, while the depth h = 6 km was modified to become 4 km. For the aftershock we adopted the value Mw = 4.5 (Cagnetti et al. (1976) give M = 4.5 without specifying the type of M), while the depth haft = 6 km was modified to become 3 km. This minimal adjustment of the depths leads to a satisfactory agreement between the isoseismals of level V (see Fig. 4). The discrepancies for the southern part of the boundary are quite understandable, since the structural parameters of a transition zone seem to be valid there.

***The earthquake of 10/10/1995***, $M_L = 5.1$, $I_0 =$ VI, ***Lunigiana*** (Ligurian Apennines, Northern Italy). The *I* map is from BM (1995), $N_{obs} = 330$, and the result of the application of the DB method is shown in Fig. 5.

The event is strike slip as indicated by the FPS parameters, dip = 80°, rake = 170°, and by the DB boundary for *I* = IV (Fig. 5a). One objection against drawing that inference from the macroseismic data consists in the fact that one lobe of the cross−shaped *I* = IV isoseismal is at sea, where no observations are available. A very good agreement for the level IV isoseismals of *I* and $I_a$ is achieved by modifying a single parameter, namely, adding +30° to the fault azimuth. That (possibly overestimated) correction affects the orientation of the isoseismal, but not its shape. The calculation of $â_p$ was for $M_w$ = 5.1 (NEIC) and $h = 2$ km (Frepoli and Amato, 1997) selected from four known depth values in the range (2–10) . The greater southward elongation of the empirical DB isoseismals of levels *I* = IV and V, compared with theory, can well be accounted for by low attenuation in zone G (Toscana).

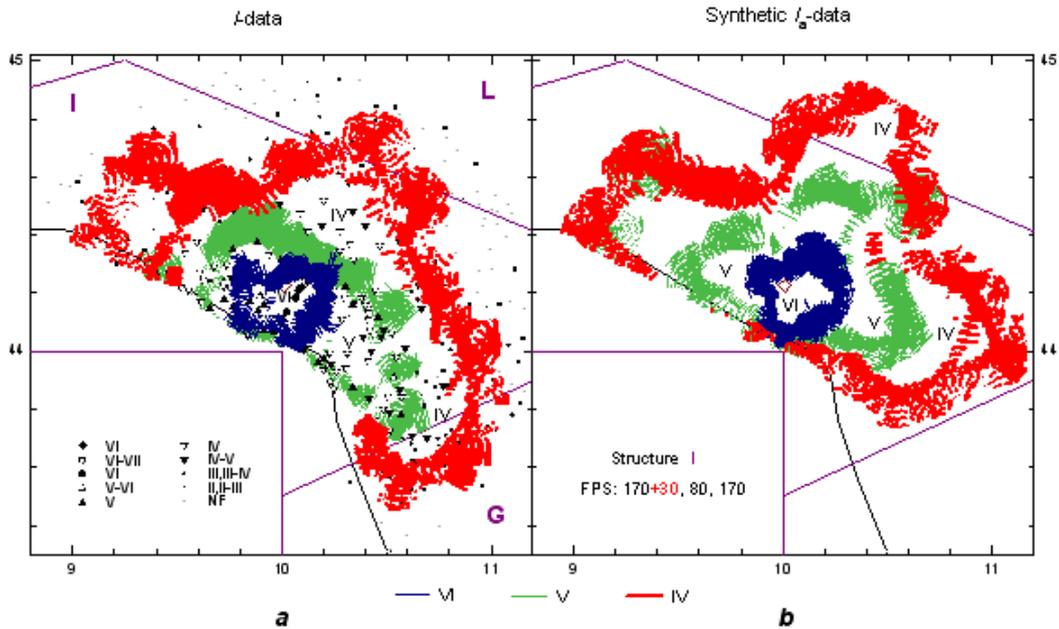

**Fig. 5**. 10/10/1995 Lunigiana earthquake: epicenter (44.23, 10.01); $h$ = 2 (Frepoli Amato, 1997), 5 (CS, 2001), 7±4 (Tertulliani Maramai, 1998), 10 (NEIC); FPS = (170°,80°, 170°) (Frepoli Amato, 1997, class A); $M_L$ = 5.1 (NEIC), 5.3 (ISC); a) raw data (symbols) and DB isoseismals of level $I$ = VI, V, IV; b) theoretical DB isoseismals for $I_a$ = VI, V, IV. Background: epicenter (rhombus) and structural zones **I**, **G**, **L**.

The event is strike slip as indicated by the FPS parameters, dip = 80°, rake = 170°, and by the DB boundary for $I$ = IV (Fig. 5a). One objection against drawing that inference from the macroseismic data consists in the fact that one lobe of the cross–shaped $I$ = IV isoseismal is at sea, where no observations are available. A very good agreement for the level IV isoseismals of $I$ and $I_a$ is achieved by modifying a single parameter, namely, adding +30° to the fault azimuth. That (possibly overestimated) correction affects the orientation of the isoseismal, but not its shape. The calculation of $â_p$ was for $M_w$ = 5.1 (NEIC) and $h$ = 2 km (Frepoli and Amato, 1997) selected from four known depth values in the range (2–10) . The greater southward elongation of the empirical DB isoseismals of levels $I$ = IV and V, compared with theory, can well be accounted for by low attenuation in zone G (Toscana).

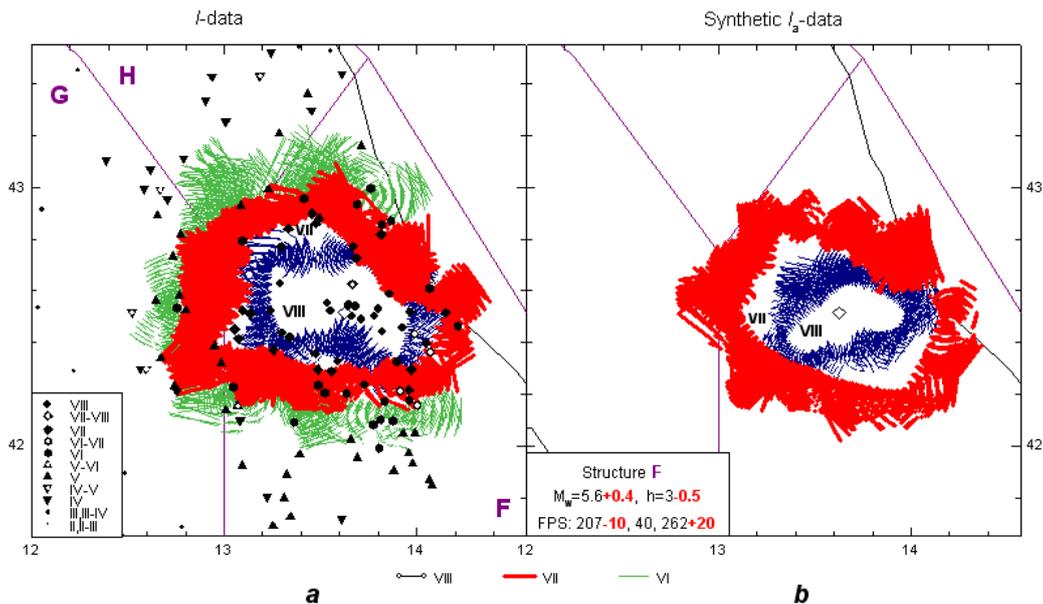

**Fig. 6**. 05/09/1950 Gran Sasso earthquake:
epicenter (42.50, 13.60); $h$ = 3 (NT, 1997), 10 (GI, 1985);FPS = (207°, 40°, 262°) (GI, 1985), $M_L$ = 5.6 (NT, 1997); a) raw data (symbols) and DB isoseismals of level $I$ = VIII, VII, VI; b) theoretical DB isoseismals for $I_a$ = VIII, VII. Background: epicenter (rhombus) and structural zones **F**, **G**, **H**.

*The earthquake of 05/09/1950*, $M_L$ = 5.6, $I_0$ = VIII, **Gran Sasso** (Abrutian Apennines, Central Italy). The $I$ map is from (Monachesi and Stucchi, 1997), $N_{obs}$ = 136, and the result of the application of the DB method is shown in Fig. 6.

In our experience, the junction of zones **G**, **H** and **F** in the central Apennines (Fig. 1) is one of the most difficult regions in Italy for isoseismal analysis. The boundaries of the structural zones deserve a special study. The $I$ = VII isoseismal (rank 1) for the Gran Sasso earthquake is mostly in the structural zone **F**, but its western part covers the boundary of zones **G** and **H**. This circumstance, in principle, should require a more sophisticated procedure than the one used here for the analysis and modeling of the $I$ = VII isoseismal. Nevertheless, adding +0.4 to $M_L$=5.6, -10° to the azimuth 207°, +20° to the rake 262°, and -0.5 km to the depth $h$ = 3 km, we have achieved a very good agreement for the isoseismal shape at level VII of $I$ and $I_a$. These variations of the parameters are quite justified for an event occurring in the mid–20th century.

*The earthquake of 15/01/1968*, $M_L$ = 6.0, $I_0$ = X, **Valle del Belice** (Sicily). The $I$ map is from Boschi et al. 2000), $N_{obs}$ = 168, and the result of the application of the DB method is shown in Fig. 7.

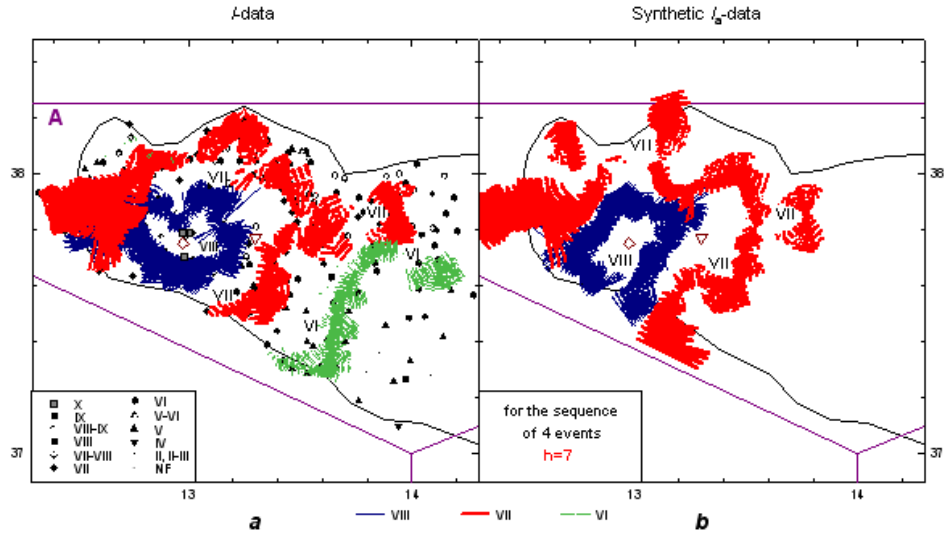

**Fig. 7.** 15/01/1968 Valle del Belice earthquake:
epicenter (37.75, 12.97); $h$ = 3 (GI, 1985), 10 (AJ, 1987), 44 (NT, 1997); FPS = (270°, 50°, 35°) (AJ, 1987), (204°, 70°, 15°) (GI, 1985); $M_L$ = 6.0 (CCI, 1997), 5.9 (NT, 1997). Foreshock: $h$ =20 (GI, 1985), 48 (CCI, 1997); FPS = (40°, 82°, 46°) (GI, 1985), $M_L$ = 5.6 (Karnik, 1969).
1–day aftershock: $h$ = 14 (GI, 1985), 36 (AJ, 1987), 47 (CCI, 1997); FPS = (250°, 58°, 18°) (AJ, 1987), (327°, 59°, 3°) (GI, 1985); $M_L$ = 5.5 (CCI, 1997), 5.6 (Karnik, 1969). 10–day aftershock: $h$ = 3 (AJ, 1987), 4 (GI, 1985), 30 (CCI, 1997); FPS = (270°, 64°, 31°) (AJ, 1987), (4°, 90°, 125°) (GI, 1985); $M_L$ = 5.7 (CCI, 1997), 5.5 (Karnik, 1969); a) raw data (symbols) and DB isoseismals of levels $I$ = VIII, VII, VI;
b) theoretical DB isoseismals for $I_a$ = VIII, VII. Background: epicenter (rhombus), epicenter of 1–day aftershock (triangle) and structural zone **A**.

The event includes a main shock with $M_L$ = 5.9–6.0, a large foreshock with $M_L$ = 5.6 which occurred 30 minutes before the main shock, and two aftershocks with $M_L$ = 5.5–5.6 and $M_L$ = 5.5–5.7 occurred 1 and 10 days after the main shock, respectively. For all the events one and the same epicenter is reported, except for the first aftershock. For the main shock two FPS solutions have been determined: (250°, 50°, 35°), (204°, 70°, 15°). It is very difficult to conduct parametric experiments when dealing with a composite event. Therefore we have limited our analysis to the following set of parameters: magnitude of the main shock, $M_w$ = 6.0, and magnitudes of the aftershocks, $M_w$ = 5.5 and 5.6, depth $h$ = 7 km for all events. The literature and catalogs report very discrepant data on the depths, giving values ranging from 3 to 44 km. The comparison of the DB isoseismals at level VII of $I$ and $I_a$ (Fig. 7) shows good agreement between theory and observations. We recall that the value of $I_a$ at a site is the largest of the theoretical values of $I_a$ corresponding to the main shock and its fore– and aftershocks.

## 5. Conclusion and Discussion

We have presented results (Figs. 2 through 7) of a comparative analysis of empirical and theoretical isoseismal shapes for six Italian $M_L$ = 5−6 earthquakes. Our modeling of isoseismals uses a plane−stratified earth model and reduces the number of parameters to a minimum. The statement applies both to our description of the source and to the method used for calibrating the theoretical intensity.

Our comparison of isoseismals is a qualitative one, but it is based in this particular case on a special method of isoseismal visualization that incorporates isoseismal uncertainty (Molchan et al., 2002). This circumstance removes from isoseismals the reputation of beeing a subjective tool for the *I*−data analysis. The examples in Figs. 2 through 7 display the shape of the lower isoseismals of level $I_{MCS}$ = IV−VII and show that their shape can be well fitted with models. In particular, Figs. 2,3,5 clearly show a radiation pattern (cross shape) related to the source geometry. In five cases of the six, the agreement in shape is achieved by varying a single parameter, namely, depth of focus or azimuth, with respect to values given in the literature. (We have not bothered to choose a suitable parameter value, when this is found in the literature in a few variants). The computational complexity does not admit a simultaneous successive variation of several parameters. Consequently, our fitting is not the result of a complete systematic variation of parameters. On the other hand, the fact that the fitting of the five events has been easy argues for the informativeness of the diffuse boundaries, *I* =IV−VII (MCS), for moderate earthquakes.

We have made a point of mentioning other 10 events (Fig. 1) where our theoretical calculations are not sufficient to substantiate the observed isoseismal shapes. During the course of this work we found that the isoseismal shapes are dependent on the earth's velocity parameters. To simplify the calculation of the wave fields we use a lateral averaging of the earth model for Italy in zones **A** through **N** (Fig. 1). Three of these (**F**, **G**, **H**) cover central Italy, which is too crude an approximation judging from the literature (see, e.g., Della Vedova et al., 1991; Chimera et al., 2002). Five of these ten events occurred in this part of Italy (Fig. 1), therefore it is not ruled out that the negative result for these five events is due to the crudeness of the velocity model.

In this connection we wish to draw the attention to those methods of *I* modeling which disregard the earth structure (Berardi et al., 1995; Sirovich, 1996). Such approaches are relevant to larger events and higher intensities. Owing to the simplicity of the calculation involved they allow practically complete successive variation of the parameters whose number is increased by 3 when an extended source is concerned (the Mach number plus source dimensions). Berardi et al. (1995) use three more parameters for the calibration of theoretical intensity, and this may result in overfitting the *I*−data.

The earthquake source information, as contained in the lower isoseismals of relatively small events ($5 \leq M_L \leq 6$) is not obvious. This is borne out by the persistent tendency of drawing isoseismals as ovals. Our examples of a fine relation between *I*−data and models are derived under at least three crude assumptions: regionalization (Fig. 1), calibration of theoretical intensity (2), and possibly the frequency range (f ≤ 1 Hz), as far as the analysis of $I \leq VI$ is concerned. This is, to some degree, a certain reserve for further analysis of events from the catalog of Kronrod et al. (2002).

*Acknowledgements.* Dr. G. Leydecker and the other, anonymous, reviewer have made many critical remarks, which have been gratefully accepted and taken into consideration. We are very grateful to A. Gusev for a discussion of our work, as well as to A. Frepoli for the materials he let us to use and for his FPS computations for the Pasubio earthquake. This work was supported by NATO SfP 972266, COFIN2000-2002 (MIUR), and by the James McDonell Foundation.

George Molchan: *e-mail*: molchan@mitp.ru;  *Tel.:* +007 (095) 333-4556